\documentstyle[aps,twocolumn,epsf]{revtex}

\newcommand{\beq}{\begin{equation}}
\newcommand{\eeq}{\end{equation}}
\newcommand{\bea}{\begin{eqnarray}}
\newcommand{\eea}{\end{eqnarray}}
\newcommand{\nn}{\nonumber}

\def\gmf{\gamma _{5}}

\def\la{\langle }
\def\ra{ \rangle }

\newcommand{\e}{{\rm e}}
\newcommand{\q}{{\bf q}}
\newcommand{\C}{{\cal C}}
\newcommand{\D}{{\cal D}}

\newcommand{\Z}{{\rm Z\!\!Z}}
\newcommand{\MM}{{\cal M}}
\newcommand{\X}{{\bf X}}

\newcommand{\junk}[1]{}

\begin{document}

\twocolumn[\hsize\textwidth\columnwidth\hsize\csname@twocolumnfalse\endcsname
\title{Coulomb Gas Representation of Low Energy $QCD$}
\author{Sebastian Jaimungal and Ariel R. Zhitnitsky}
\address{Department of Physics and Astronomy, University of British Columbia,
Vancouver, BC V6T 1Z1, Canada}
\date{\today}
\maketitle
\begin{abstract}
A novel Coulomb gas (CG) description of low energy $QCD_4$, based on
the dual transformation of the QCD effective chiral Lagrangian, is
constructed.  The CG is found to contain several species of charges,
one of which is fractionally charged and can be interpreted as
instanton-quarks. The creation operator which inserts a
pseudo-particle in the CG picture is explicitly constructed and
demonstrated to have a non-zero vacuum expectation value. The Wilson
loop operator as well as the creation operator for the domain wall in
the CG representation is also constructed.
\end{abstract}
\vskip 2mm] 

{\bf 1.} Color confinement, spontaneous breaking of chiral symmetry,
the $U(1)$ problem, $\theta$ dependence, and the classification of
vacuum states are some of the most interesting questions in
$QCD$. Unfortunately, the progress in our understanding of them is
extremely slow. At the end of the 1970s A. M. Polyakov \cite{Po77}
demonstrated color confinement in $QED_3$; this was the first example
in which nontrivial dynamics was a key ingredient in the solution.
Soon after, 't Hooft and Mandelstam \cite{Hooft} suggested a
qualitative picture of how confinement could occur in $QCD_4$. The key
point, the 't Hooft - Mandelstam approach, is the assumption that
dynamical monopoles exist and Bose condense. Many papers have been
written on this subject since the original formulation \cite{Hooft};
however, the main questions, such as, ``What are these monopoles?'';
``How do they appear in the gauge theories without Higgs fields?'';
``How do they interact?'', 
``What is the relation (if any) between confinement
and instantons?"
were still not understood. Almost 20 years
passed before the next important piece of the puzzle was solved
\cite{SeWi94}. Seiberg and Witten demonstrated that confinement occurs
in SUSY $QCD_4$ due to the condensation of monopoles much along the
lines suggested many years ago by 't Hooft and Mandelstam (for a
recent review see \cite{Hooft1}). Furthermore, condensation of dyons
together with oblique confinement for nonzero vacuum angle, $\theta$,
was also discovered in SUSY models (a phenomenon which was also argued
to take place in ordinary $QCD$; see \cite{Hooft1}). In addition to
forming concrete realizations of earlier ideas, the recent progress in
SUSY models has introduced many new phenomena, such as the existence
of domain walls\cite{Shifman1} which connect two distinct vacua with
the same $\theta$.
%  
%New insights into confinement was also recently
%given by Witten \cite{Wi98}, in which he argued that domain walls
%connecting two vacua labeled by $k$ and $k+1$ behave similar to
%$D$-branes on which the $SQCD$ strings can end. It is tantalizing to
%suggest that such phenomenon also take place in $QCD$; indeed, in this
%letter it is argued that this in fact does occur.
%

In this letter we shall illuminate one of the missing elements
mentioned above by demonstrating that the dual representation of the
low-energy effective chiral Lagrangian corresponds to a statistical
system of interacting pseudo-particles with fractional $1/N_c$
charges.  We shall identify these particles with instanton quarks
suspected long ago\cite{Fateev}, \cite{Belavin}, consequently
demonstrating a link between confinement and instantons.

Through a rewriting of the low energy effective $QCD_4$ action
\cite{HZ}, in terms of a dual Coulomb gas (CG) picture, the operator
which creates pseudo-particles will be demonstrated to have a non-zero
vacuum expectation value (vev).  Many additional interesting features
will arise in the dual CG picture:
\\[0mm] \indent i) there are several
species of pseudo-particles which interact according to the Coulomb
law;
%
% in a manner reminiscent of the
%'t~Hooft ``determinant'' interaction;
%
\\[0mm] \indent ii) the species
conjugate to the flavor singlet component of the chiral condensate
will be shown to correspond to fractionally charged pseudo-particles;
\\[0mm] \indent iii) the operator which creates the pseudo-particles
in the dual CG picture will be constructed in terms of the phases of
chiral condensate; \\[0mm] 
%
%\indent iv) the existence of almost
%degenerate vacua lead to non-trivial classical (unstable) solutions;
%and the interactions between the pseudo-particles in the dual picture
%with these ``domain walls'' are identified. Within this context a
%connection with Witten's \cite{Wi98} interpretation of these ``walls''
%as $D$-branes, on which the $QCD$ string can end, will be made;
%\\[0mm]
%
\indent iv) the Wilson loop operator insertion in the dual CG
picture will also be constructed.

Our analysis begins with the effective low energy $QCD$ action derived
in \cite{HZ}, which allows the $\theta$-dependence of the ground state
to be analyzed. Within this approach, the Goldstone fields are
described by the unitary matrix $U_{ij}$, which correspond to the
$\gmf$ phases of the chiral condensate: $ \la \overline{\Psi}_{L}^{i}
\Psi_{R}^{j} \ra = - | \la \overline{\Psi}_{L} \Psi_{R} \ra | \,
U_{ij}$ with $U = \exp \left[ i \sqrt{2} \, \frac{\pi^{a} \lambda^{a}
}{f_{\pi}} + i \frac{ 2}{ \sqrt{N_{f}} } \frac{ \eta'}{ f_{\eta'}}
\right] , \quad U U^{+} = 1,~ $where $ \lambda^a $ are the Gell-Mann
matrices of $ SU(N_f) $, $ \pi^a $ is the pseudo-scalar octet, and $
f_{\pi} = 133 \; MeV $.  In terms of $U$, the low-energy effective
potential is given by\cite{HZ}: 
\bea W_{QCD} &=&- \lim_{V
\rightarrow \infty}~ \frac{1}{V} ~ \ln \sum_{r=-\infty}^{\infty}
\sum_{l=0}^{p-1} ~\exp \Big\{\nn\\ && V E \cos \left(- \frac{q}{p}
\theta + i\frac{q}{p} \log {\rm Det}~ U + \frac{2 \pi}{p}~l - 2\pi r
\right) \nn\\ && +\frac{1}{2} V \, {\rm Tr} ~( m U + m^{+} U^{+} )
\Big\},
\label{2} 
\eea 
where $V$ is the volume of the system.  All
dimensional parameters in this potential are expressed in terms of the
$QCD$ vacuum condensates, and are well known numerically: $ m = {\rm
diag} (m_{q}^{i} | \la \overline{\Psi}^{i} \Psi^{i} \ra | )$; and the
constant $E$ is related to the $QCD$ gluon condensate $ E = \la b
\alpha_s /(32 \pi) G^2 \ra $. The only unknown parameters in this
construction are the integers $p,q$, which play the same role as the
discrete integer numbers classifying the vacuum states in SUSY
theories. The only constraint on them is that in large $N_c$ limit,
$q/p\sim 1/N_c$ such that the $U(1)$ problem is resolved.

To convince the reader that (\ref{2}) does indeed represent the
anomalous effective Lagrangian, three of its most salient features are
listed below (for details see \cite{HZ}):\\[0mm]
\indent i)
Eq.. (\ref{2}) correctly reproduces the VVW effective chiral Lagrangian
\cite{Wit2} in the large $ N_c $ limit; \\[0mm]
\indent ii)
it reproduces the anomalous conformal and chiral Ward identities of
$QCD$;\\[0mm]
\indent iii) 
it reproduces the known dependence in $\theta$ for small angles
\cite{Wit2}; however, it may lead to different behavior for large values
$\theta >\pi/q$ if $q\neq 1$. Accordingly, it leads to the correct $2\pi$
periodicity of observables.  

From this point onwards  $U=\exp\{i{\rm
diag}(\phi_1,\dots,\phi_{N_f})\}$, are the phases of the chiral
condensate and $\phi = {\rm Tr~ln}~U$ represents the singlet 
field\footnote{ It is well known that this is the
most general form for $U$-matrix describing the ground state of the
system\cite{Wit2}. The off-diagonal elements of $U$ describe the
fluctuations of the physical Goldstones which are neglected. This is
allowed since only the diagonal elements are relevant in the
description of the ground state which is the focus of this
work. Furthermore, such a truncation can be justified a posteriori by
demonstrating that the classification of vacuum states based on the CG
representation exactly coincides with the classification based on the
effective Lagrangian approach, see Eq. (\ref{2}) where only the
diagonal elements are relevant.  Finally, as will be demonstrated
shortly, the most important contribution is related to the {\it
singlet field}, $\phi = {\rm Tr~ln}~U$, which is unambiguously defined.
In fact, all contributions related to the non-singlet fields are
suppressed in the chiral limit by the quark masses $m_i$, and can be
neglected as a first approximation.}. The
effective potential then takes on a specific Sine-Gordon (SG) form.
Such a structure is quite natural for terms proportional to $m_i$ and
is associated with the Goldstone origin of the $\phi_i$ fields.  A
similar Sine-Gordon structure for the singlet combination is less
obvious and corresponds to the following behavior of the $(2k)^{\rm
th}$ derivative of the vacuum energy in pure gluodynamics, $$ \left.
\frac{ \partial^{2k} E_{vac}(\theta)}{ \partial \, \theta^{2k}}
\right|_{\theta=0} \sim \int \prod_{i=1}^{2k} dx_i \la
Q(x_1)...Q(x_{2k})\ra \sim (\frac{i}{N_c})^{2k},
$$ where $Q\sim G_{\mu\nu} {\widetilde G}_{\mu\nu}$. This property was
seen as a consequence of the solution of the $U(1)$ problem
\cite{Venez}.
 
{\bf 2.} Although there are several interacting fields in the
effective $QCD$ action (\ref{2}) many of the special properties of the
SG theory apply to this model, and the admittance of a CG
representation \cite{Po77} for the partition function is no different.
The existence of many fields and cosine terms only serve to make the
formulae more bulky while the basic strategy remains the same.  Using
the effective potential (\ref{2}) the defining partition function is
taken to be\footnote{The following fact was used to replace the
summations over $r$ and $l$ in (\ref{2}) with $k$ and $n$ in
(\ref{Z}): given the pair of integers $(r \in \Z,l = 0,...,p-1)$ there
exists a unique pair of integers $(n \in \Z, k = 0,...,p-1)$, such that:
$\frac{q}{p} n - k = \frac{l}{p} - r$}, \bea Z&=& \sum_{k=0}^{p-1}
\sum_{n=-\infty}^{\infty} \int \D \phi_a~ \e^{- g \int d^4 x ({\vec
\nabla} \phi_a)^2} \label{Z}\\ && \times \;e^{\int d^4 x \Big\{ E \cos
\left( \frac{q}{p} \left(\phi - \theta + 2\pi n \right) + 2\pi k
\right) + \sum_{a=1}^{N_f} m_a \cos (\phi_a) \Big \} }\nn \eea where
the coupling constant $g \propto f_\pi^2$.  Performing a series
expansion in $E$ and $m_a$, and introducing the $\Z_2$ valued fields,
$Q^{(a)}$ for $a=0,\dots,N_f$, to represent the cosine interactions,
makes it possible to rewrite the partition function in a form in which
the dynamical scalar fields can be integrated out exactly. Carrying
out the integration then leads to the novel dual CG picture with
action, 
\bea S_{CG} &=& i \theta Q^{(0)}_T - \ln \left(p
\sum_{n=-\infty}^{\infty} {\rm e}^{2\pi i Q^{(0)}_T n} \right) \nn\\
&& + \frac{1}{2g^2} \sum_{a=1}^{N_f} \Bigg\{ \sum_{b,c=1}^{M_0}
Q^{(0)}_b~G(x^{(0)}_b-x^{(0)}_c)~Q^{(0)}_c \nn\\ && \hspace{14mm} + 2
\sum_{b=1}^{M_0} \sum_{c=1}^{M_a} Q^{(0)}_b~G(x^{(0)}_b-x^{(a)}_c)
~Q^{(a)}_c \nn\\ && \hspace{15mm} + \sum_{b,c=1}^{M_a}
Q^{(a)}_b~G(x^{(a)}_b-x^{(a)}_c)~Q^{(a)}_c \Bigg\}. 
\label{CGaction}
\eea 
The species $Q^{(0)}_i$ are dual to the singlet field $\phi(x)$,
while the species $Q^{(a\ne0)}_i$ are dual to the phases, $\phi^a(x)$,
of the chiral condensates $\langle
\overline{\Psi}^a\Psi^a\rangle$. Also, $Q^{(0)}_T=\sum_{b=1}^{M_0}
Q^{(0)}_b$ is the total $Q^{(0)}$ charge for that configuration and
$G(x-y)$ denotes the relevant Greens function of the Laplace
operator. The full partition function is, \bea Z =
\sum_{M_0,\dots,M_{N_f}=0}^{\infty} \sum_{\stackrel{\scriptstyle
Q^{(0)}_i = \pm \frac{q}{p}} {Q^{(a\ne0)}_i=\pm 1}}\!
\frac{(\frac{E}{2})^{M_0}}{M_0!}\frac{(\frac{m_1}{2})^{M_1}}{M_1!}
\dots\frac{(\frac{m_{N_f}}{2})^{M_{N_f}}}{M_{N_f}!} \nn \eea \bea
\times \int (dx_{1}^{(0)} \dots d x_{M_0}^{(0)})
\dots(dx_{1}^{(N_f)}\dots d x_{M_{N_f}}^{(N_f)}) ~ \e^{-S_{CG}}.
\label{CG} 
\eea Notice that the fugacities of the species-$(a\ne0)$
are given by the masses of the $a^{\rm th}$ quark, while the fugacity
of the species-$(0)$ is proportional to the gluon
condensate $E$.  An important point is that the fugacity of the
species-$(a\ne0)$ vanishes in the chiral limit, while that of the
$(0)$ species remains non-zero. A second point is that, in the
chiral limit, it is not obvious whether (\ref{CG}) is independent of
$\theta$, while in $QCD$ the $\theta$-angle appears with quark masses
and hence disappears in this limit. It is possible to demonstrate that
the partition function has this property; however, the details are
deferred to an extended version of this letter \cite{JaZh}.

{\bf 3.} There are several important features of the action
(\ref{CGaction}) which should be noted. Firstly, the summation over
$n$ forces the total $Q^{(0)}$ charge, $Q_T^{(0)}$, to be an
integer. Such a constraint is the analog of the quantization of the
topological charge and is to be expected.  Secondly, due to the manner
in which $Q^{(0)}$ appears with the parameter $\theta$ and the fact
that its total charge, $Q_T^{(0)}$, is integer, one can identify
$Q_T^{(0)}$ as the total topological charge (defined in four
dimensional Euclidean space) of the given configuration.
%
% Secondly, 
%the charge $Q^{(0)}$ was originally
%introduced in a very formal manner so that the 
%QCD effective low energy Lagrangian can be written in the dual
%CG form (\ref{CGaction}). The physical meaning of these
%charges was not obvious at the time when they have been introduced.
%Now, the physical 
%  interpretation of these charges becomes clear: since $Q_T^{(0)}$
%is an integer number(charge) which appears in the action $S_{CG}$
% together with the parameter $\theta$  (\ref{CGaction}), one could deduce that
%$Q_T^{(0)}$ is nothing but the total topological charge 
%(defined in the  four dimensional
%Euclidean space) of the given configuration. 
%
Indeed, in QCD the $\theta$ parameter appears in the Lagrangian only
in the combination $i\theta\frac{ G_{\mu\nu} {\widetilde
G}_{\mu\nu}}{32\pi^2}$.  It is quite interesting that although the
starting effective low-energy Lagrangian is colorless, we have ended
up with a representation in which a colorfull object (the topological
charge $Q_T^{(0)}$) can be identified.
%
%It is quite amusing result: we
%started from the effective low-energy Lagrangian which is clear the
%colorless object; we end up with CG representation where $Q_T^{(0)}$
%is identified with the total topological charge which is obviously a
%description of a colorful object.??
%
Furthermore, since the $(0)$ species has charges $\sim\frac{q}{p}\sim
1/N_c$, the integer constraint enforces a fractional quantization on
the pseudo-particles: the difference in the number of positively and
negatively charged pseudo-particles of species-$(0)$ must be an
integer multiple of $p$.  This simple observation implies that our
fractionally charged pseudo-particles, $Q^{(0)}$, cannot be related to
any semi-classical solutions, which can carry only integer charges;
rather, configurations with fractional charges should have pure
quantum origin. 

 Let us reiterate the reasons for this identification:
The fact that species $Q^{(0)}_i$ has charges $ \sim 1/N_c$ is a
direct consequence of the $\theta/N_c$ dependence in the underlying
QCD with frozen, non-dynamical quarks; in addition, due to the $2\pi$
periodicity of the theory, only configurations which contain an integer
topological number contribute to the partition function. Therefore, 
the number of particles $Q^{(0)}_i$ with charges $ \sim 1/N_c$ 
must be proportional to $N_c$.

Note that the $\theta$-angle only appears with $Q^{(0)}_T$, this is a
direct consequence of the pseudo-particles of species-$(0)$ being dual
to the singlet field $\phi$ which is the only field appearing with
$\theta$ in the effective low-energy action (\ref{2}).
%
%a nonzero contribution to the
%partition function comes only from the configurations with an integer
%number $\sim N_C$ of pseudo-particles.??  Thirdly, the
%$\theta$-angle only appears with the $(0)$ species. This is a result
%of species-$(0)$ being dual to the singlet field $\phi$, which is the
%only field that is directly affected by the $\theta$-angle. 
%
A final point about the form of the CG action is that the
$\theta$-dependence supplies an overall phase factor for each
configuration and leads to the very natural interpretation of
non-trivial $\theta$-angles as introducing an overall background
charge. Turning attention to the interactions amongst the
pseudo-particles: species-$(0)$ is seen to interact with all species-($a=0,\dots,N_f$); however, all other species, $(a\ne0)$, only
interact with their own species and species-$(0)$. This peculiar
behave is, once again, due to species-$(0)$'s association with the
singlet while the other species are associated with particular
components of the chiral condensate.
%
%It is interesting that the
%structure of interactions between the pseudo-particles resembles the
%'t~Hooft ``determinant'' interaction. {\bf more explanation about
%'t~Hooft determinant}.  Although the interactions appearing in
%(\ref{CG}) are two-body, species-$(0)$ interacts with all other
%species, allowing for vertices with $1,2,3,..., N_f$ legs.  This is
%illustrated in figure \ref{fig1}.
%\begin{figure}[t]
%\epsfxsize = 8.5cm
%\epsffile{interactions.eps}
%\caption{Each species interacts with itself (left most diagram), and with 
%species $0$. Notice that species $0$ can interact with up to $N_f$
%distinct species of pseudo-particles.\label{fig1}} 
%\end{figure}
%

In order to attach a physical meaning to each charge $Q^{(0)}_i$, as
opposed to the total topological charge $Q_T^{(0)}$, we would like to
remind the reader about an interesting connection between the CG
statistical ensemble and the $2d$ $O(3)~\sigma$-model (more generally,
$CP^{N_c}$ models) \cite{Fateev}. An exact accounting and resummation
of the $n$-instanton solutions maps the original problem to a $2d$-CG
with fractional charges (the so-called instanton-quarks).  These
pseudo-particles, instanton-quarks, do not exist separately as
individual objects; rather, they appear in the system all together as
a set of $\sim N_c$ instanton-quarks so that the total topological
charge of each configuration is always integer. This means that a
charge for an individual instanton quark cannot be created
and measured;
instead, only the total topological charge for the whole configuration
is forced to be integer and has a physical meaning.
  However, a convenient parameterization of the
instantons can be made by associating a charge of strength $\sim
1/N_c$ to the pseudo-particles. In fact, such an interpretation in
terms of these (fictitious) fractionally charged objects leads to an
elegant explanation of confinement and other important properties of
the model \cite{Fateev}.
%
%one should say, that this parameterization is so convenient that the
%confinement and other important properties of the model can be easily
%explained in terms of these (fictitious) fractionally charged
%objects\cite{Fateev}. ??
%
With this in mind, the discussion is now returned to the CG system
(\ref{CGaction}). Notice that if $q=1$ and $p=N_c$ (as arguments based
on SUSY suggest \cite{Shifman1}), then, the number of integrations
over $d^4x_i^{(0)}$ in Eq.(\ref{CG}) exactly equals $4 N_c k$, where $
k$ is integer. This is a consequence of the fact that the number of
species-$(0)$ pseudo-particles must be an integer multiple of
$N_c$, see
discussion after Eq.(\ref{CG}). This number, $4 N_c k$, exactly
corresponds to the number of zero modes in the $k$-instanton
background, and we {\it conjecture} that (at low energies\footnote{The
analysis carried out here is based upon a low-energy effective action,
as such nothing can be said about the high energy behavior of our
dual model.}) the fractionally charged species-$(0)$ pseudo-particles
are the instanton-quarks suspected long ago\cite{Belavin}.  An
interesting point is that if the gauge group, $G$, was not $SU(N)$,
the number of integrations would be equal to $4k C_2(G)$ where
$C_2(G)$ is the quadratic Casimir of the gauge group. This is because
the $\theta$ dependence in physical observables will then appear in
the combination $\frac{\theta}{C_2(G)}$, and all appearances of $N_c$
are replaced by $C_2(G)$.  This number $4k C_2(G)$ exactly corresponds
to the number of zero modes in the $k$-instanton background for gauge
group; hence, this analysis correctly reproduces the scenario for
arbitrary gauge groups, and supplies further support for our
conjecture.
%
%We should
%also point out that an $SU(N)$ gauge group is not special in this
%respect. Analogous, very unexpected relation between number of
%integration over $d^4x_i^{(0)}$ in Eq.(\ref{CG}) and number of zero
%modes in the $k$-instanton background, $4 C_2(G) k$ with $ C_2(G)$ as
%the quadratic Casimir operator also takes place for arbitrary gauge
%group $G$. The number $ C_2(G)$ emerges in this formula because the
%$\theta$ dependence appears in the physical observables in the
%combination $\frac{\theta}{ C_2(G)}$.

{\bf 4.} The relationship between a CG representation for the
classical gas of pseudo-particles and the effective quantum field
theory for a scalar field $\phi$ is not a new idea and has been known
since\cite{Po77}. The general strategy in this approach is to express
all observables (Wilson loop, etc.)  in terms of the gas of
pseudo-particles and use the effective description in terms of the
auxiliary $\phi$ field simply for convenience of calculations.
However, it has proven to be useful\cite{Snyderman} to consider the
$\phi$ field as a main player and to obtain the operator in the $\phi$
description which introduces a pseudo-particle at the point $x$ in the
CG (this operator will be denoted $\MM$).  The analysis
\cite{Snyderman} of the corresponding correlation functions in the
Polyakov model of the operator $\MM$ leads to the same explanation of
confinement which was already known. However, new insights about
confinement extending from such analysis (e.g. condensation being a
consequence of $\la\MM\ra\neq0$) are found. In particular, the $\phi$
field has the physical meaning of a potential (indeed, the action
density of the system is proportional to $-\cos(\phi)$) describing a
statistical ensemble of pseudo-particles in the mean-field
approximation.  The condensate $\la\MM\ra \sim e^{i\la\phi\ra}\neq 0$
was named ``magnetization'', and the $\phi$ field was called the scalar
magnetic potential due to the fact that pseudo-particles in 3d are
magnetic monopoles.

We would like to carry out a similar analysis, and construct the
operator, $\MM$, which creates a pseudo-particle (instanton quark) at
the point $x$.
%
%We want to go along this road and construct the creation
%operator $\MM$ of a pseudo-particle (instanton quark) at point $x$.??
%
It can be demonstrated \cite{JaZh} that the operator which inserts a
single pseudo-particle of species-$(a)$ with charge, $q_a$, in the
bulk of the gas can be written in terms of the phases of the chiral
condensate as follows: 
\beq
\label{10}
\MM(\q_a,\X) = \left\{ 
\begin{array}{ll}
{\rm e}^{i\q_0\left( 
\phi(\X) - \theta+2\pi n + 2\pi \frac{p}{q} k\right)} &,\quad a = 0 \\
 {\rm e}^{ i \q_a\phi_a(\X)} &, \quad a\ne 0 
\end{array}
\right.
\label{Q0insert}
\eeq Such an exponential form for the creation operator $\MM$ in terms
of the scalar $\phi$ field is quite similar to what was found for
Polyakov's model\cite{Snyderman}.  This should not be surprising,
since such a form is a universal property for CG representations and does
not depend on the dimension of the space-time.

With equal, small, quark masses, $m\rightarrow 0$, the vev's in the
semi-classical limit are $\langle \phi_a\rangle\sim
\frac{\theta}{N_f}$ for small $\theta$ \cite{HZ} implying that in this
regime, $\langle \MM \rangle \ne 0$.  Recall that our starting point,
the effective action, was constructed under the assumption that the
system is in the confining phase, i.e. (\ref{2}) contains only
colorless degrees of freedom.   This confinement
(which was already implemented into our system) is described
 in terms of the $\phi_a$ fields  
by a non-zero vev $\langle \MM \rangle \sim {\rm e}^{ i \la
\q_a\phi_a\ra } \neq 0 $.  Therefore, it is tempting to assume that
the property $\langle \MM \rangle \ne 0$ is intimately related to
confinement in any dimension as was checked in 2d $CP^{N-1}$
 model\cite{Fateev}, 3d
Polyakov's model\cite{Snyderman} and 4d QCD(\ref{Q0insert}).

Due to the existence of non-trivial solutions to the classical
equations of motion of the effective action\cite{FHZ}, an operator in
the CG representation which induces a source for those solutions
should exist. Indeed, it is not difficult to show\cite{JaZh} that the
operator is, 
\beq \C(\eta_a) \propto {\rm e}^{i\sum_{b=1}^{M_0}
Q^{(0)}_b \eta(x^{(0)}_b)} \prod_{a=1}^{N_f} {\rm
e}^{i\sum_{b=1}^{M_a} Q^{(a)}_b \eta_a(x^{(a)}_b)}. \label{bkfield}
\eeq
where, $\eta \equiv \eta_1+\dots+\eta_{N_f}$. This operator
inserts sources, $\int d^4x \phi_a \Box \eta_a$, into the effective
action (\ref{Z}).  Consequently, fluctuations about a classical field
configuration can be taken into account in the CG picture by
introducing a phase factor which gives rise to a source term for that
background - all interactions among the charges remain unaltered.  An
interesting source is given by $\eta(x) = \frac{ 2\pi}{q} {\rm
sgn}(x_0)$ which, in the SG representation, gives rise to the domain
wall solutions that interpolate between vacuum states labeled by $k$
and $k+1$ \cite{FHZ}. Such a source term in the CG clearly interacts
non-trivially with {\it all} charges.  Similar domain walls are known
to exist in supersymmetric models (see \cite{Shifman1} for a review).
% Witten recently conjectured
%\cite{Wi98} that in the large $N_c$ limit the domain walls connecting
%vacua labeled by $k$ and $k+1$ appear to be objects which are not
%solitons from the string viewpoint; rather, they appear to be
%$D$-branes on which the string can end.  Our CG picture suggests that
%such phenomenon could take place in QCD. Indeed, a distinguishable
%property of $D$-branes, in the large $N_c$ limit, is that the tension
%behaves like $\sim N_c$. The domain walls described in \cite{FHZ} have
%precisely this property. Moreover, the pseudo-particles 
%have been demonstrated to carry fractional charges $ \sim
%\frac{1}{N_c}$.  Therefore, a fractional, $\frac{1}{N_c}$,
%chromo-electric flux carried by an open string of the corresponding
%(non-critical) string theory can end on the QCD $D$-brane.

As a final important identification, consider an insertion of the
source 
\bea
\label{eta}
   \Box\eta=4\pi \theta_{S}(x_1, x_2) 
\int dy_{\lambda}\epsilon^{\lambda\sigma}
\partial_{\sigma } \delta^2(x-y),~ \lambda,\sigma=0,3,
\eea
where $\theta_S(x_1,x_2)$ is unity within a surface $S$ and zero
outside.  One can show \cite{JaZh} that it corresponds to the
insertion of a Wilson loop operator in the $(x_1,x_2)$-plane. In the
SG picture (\ref{Z}) the Wilson loop insertion corresponds to the
shift $\phi \rightarrow\phi+\eta $ in the flavor-singlet potential
term $\sim E \cos(\frac{q}{p}(\phi+\eta-\theta +2\pi n) +2\pi k))$.

{\bf 5.}  
To conclude: the existence of the free parameter $\theta$ plays the
role of a messenger between colorless and colorful degrees of freedom.
The CG picture developed has a rich structure in which several species
of pseudo-particles appear and the unique fractionally charged
species is found to be explicitly affected by the presence of a
$\theta$-angle. 

As a final concluding remark, at the intuitive level there seems
 to be a close relation between out CG representation in terms of 
instanton quarks and the ``periodic instanton" analysis\cite{vanbaal}.
Indeed, in\cite{vanbaal}  has been shown that the large size
instantons and monopoles are intimately connected
and, therefore, instantons have the internal structure. Unfortunately,
one should not expect to be able to account
for large instantons using semiclassical technique   
to bring this intuitive correspondense onto the quantitative level.

 Besides that, the recent 
analysis\cite{SYM}, where it was demonstrated that in SYM
the instanton quarks carry magntetic charges and saturate
the gluino condensate,  also supports our picture. Finally, 
the recent analysis\cite{konishi}of dynamical symmetry
breaking in  $SU(N_c)$ and $USp(2N_c)$ gauge theories,
shows the condensation of `` magnetic quarks" with fractional
charges rather than magnetic monopoles itself. 
This phenomenon again supports our picture where we claim
that the instanton constituents (instanton quarks) are appropriate
degrees of freedom in description of low energy QCD.

Discussions 
on dense instanton ensemble 
and monopoles with Pierre van Baal, T. Schafer and E. Shuryak  
 are greatly appreciated.

\end{document}